\documentclass[10pt,a4paper]{article}

\usepackage{caption}
\usepackage{subcaption}
\usepackage{graphicx}
\usepackage{booktabs}
\usepackage{xcolor}
\usepackage{array}
\graphicspath{{figures/}}

\usepackage{amssymb}
\usepackage{amsmath} % Required for some math elements 

\usepackage{algorithmic}
\usepackage{array}
\usepackage{lipsum}
\usepackage{hyperref}

\usepackage[sort&compress]{natbib}
\begin{document}
\date{}
\title{Joint Training on AMD and NVIDIA GPUs} % Title
\author{
  Jon Hu, Thomas Jia, Jing Zhu, Zhendong Yu \\
  Zettabyte AI, Inc. \\
  \texttt{\{jonh, tomj, jing.zhu, zhendong.yu\}@zettabyte.space}
}

\maketitle % Inserts the title, author and date

%----------------------------------------------------------------------------------------
%	Abstract
%----------------------------------------------------------------------------------------
\begin{abstract}
\normalsize
%% Text of the abstract
As large language models continue to scale, training demands on compute and system capacity grow rapidly, making single-vendor homogeneous clusters insufficient. This paper presents a technical solution for heterogeneous mixed training in AMD–NVIDIA environments. We first adopt a compatibility-oriented approach based on CPU-Forwarding Communication, with differentiated communication backend selection across parallel groups and multi-NIC parallel data transfer. To achieve higher performance, we further propose another Device-Direct Communication approach, integrating a CPU-offloading P2P mechanism to enable direct cross-vendor GPU data transfer without host-memory staging. Experiments on LLaMA-8B and Qwen2-7B demonstrate that the proposed Device-Direct Communication approach achieves up to 98\% of the throughput of an NVIDIA homogeneous system, while preserving training stability and correctness.

\end{abstract}

%----------------------------------------------------------------------------------------
%	Table of Content
%----------------------------------------------------------------------------------------
\setcounter{tocdepth}{2}

%----------------------------------------------------------------------------------------
%	Appendix
%----------------------------------------------------------------------------------------
\section{Introduction}
\label{intro}

As the parameter scale of large language models (LLMs) surpasses the trillion-parameter level, computational demand increases exponentially in accordance with scaling laws \cite{kaplan2020scaling}, rendering single homogeneous clusters insufficient to meet the growing training requirements. Constrained by shortened hardware iteration cycles and increasing supply-chain diversification, modern computing clusters have become inherently heterogeneous \cite{ning2023supply}. Such heterogeneity not only arises from the coexistence of accelerators across different vendors and generations (e.g., NVIDIA \cite{h200} and AMD \cite{mi325x}), but also manifests in substantial disparities in memory capacity and interconnect bandwidth (e.g., variations in HBM capacity and interconnect technologies such as NVLink \cite{nvlink} and PCIe \cite{pcie}). Effectively aggregating these heterogeneous resources to support large-scale model training has become a critical challenge in overcoming computational bottlenecks.

To advance heterogeneous mixed training, we propose and implement an end-to-end technical solution for AMD–NVIDIA heterogeneous training. We initially adopt a compatibility-oriented approach based on CPU-Forwarding Communication, with optimizations including differentiated communication backend selection across parallel groups and multi-NIC parallel data transfer for Pipeline Parallel (PP) groups. While this enables cross-vendor interoperability, frequent host–device (H2D/D2H) data transfers prevent fully exploiting heterogeneous hardware performance. To address this limitation, we further propose a Device-Direct Communication approach, integrating GPUDirect RDMA \cite{gao2023high} (GDR) with a CPU-offloading P2P communication scheme, enabling direct data transfers between GPUs from different vendors and completely eliminating the high-overhead intermediate host-side memory copies.

Experimental results show that, when training LLaMA-8B \cite{llama} and Qwen2-7B \cite{qwen2} on an AMD–NVIDIA heterogeneous cluster, the proposed Device-Direct Communication approach achieves up to 98\% of the throughput of an NVIDIA homogeneous system, while maintaining strong stability and correctness. These results demonstrate a practical and effective pathway for efficiently leveraging diverse computational resources.

The remainder of this paper is organized as follows. Section~\ref{sec-approach} introduces two mixed-training approaches, namely CPU-Forwarding Communication and Device-Direct Communication. Section~\ref{result} presents the experimental results and analysis. Section~\ref{discussion} discusses key design choices and engineering considerations in the implementation. Finally, Section~\ref{sec:conclusion} concludes the paper.
\section{Approach}
\label{sec-approach}
\subsection{CPU-Forwarding Communication}
CPU-Forwarding Communication is a common and straightforward solution for cross-vendor GPU communication, offering the inherent advantage of abstracting compatibility barriers between disparate hardware ecosystems. Gloo \cite{gloo} is a representative CPU-side collective communication library (CCL), capable of facilitating communication across GPUs with different architectures through GPU forwarding. In Megatron framework \cite{shoeybi2019megatron}, configuring the communication backend to Gloo provides native support for heterogeneous mixed training. However, this approach relies on global GPU data being routed through the CPU, which suffers from significant performance degradation due to frequent host-device memory copies.

To achieve higher performance, we propose the following optimization strategies:

\begin{enumerate}
    \item \textbf{Differentiated Communication Backend Selection (DCBS)}: In LLM training, inter-group communication is typically low for Pipeline Parallel (PP) groups, but high for Data Parallel (DP) and Tensor Parallel (TP) groups. Consequently, we employ the Gloo backend for inter-group PP communication to support heterogeneous interconnects. In contrast, DP and TP groups remain homogeneous, continuing to leverage vendor-specific libraries (e.g., NCCL \cite{hu2025demystifying}, RCCL \cite{rccl}) to maintain peak communication performance.
    
    \item \textbf{Multi-NIC Parallel Data Transfer (MPDT)}: For cross-node PP communication, an independent Network Interface Card (NIC) is allocated to each GPU for Point-to-Point (P2P) data transfers. This facilitates the physical parallelization of PP traffic, effectively enhancing bandwidth utilization and overall training throughput.
\end{enumerate}

Implementing these strategies involves targeted modifications to the Megatron framework, specifically: the differentiated initialization of communication group backends and the implementation of explicit data copy logic between GPU and CPU buffers during P2P operations.

While the strategies above offer substantial performance gains over the native Megatron-Gloo implementation, this approach' s efficiency remains constrained by the overhead of host–device memory copies, which prevents the full realization of the computational and communicative potential of hardware.
\subsection{Device-Direct Communication}
Device-Direct Communication aims to eliminate the CPU-mediated overhead inherent in heterogeneous GPU communication. This approach facilitates direct data transfer between cross-vendor GPUs, effectively bypassing inefficient host-side memory copies. The primary challenge in realizing this path involves bridging the hardware ecosystem divide, necessitating both direct memory access via GDR and cross-library synchronization across vendor-specific CCLs.

\begin{figure}[t]
    \centering
    \begin{minipage}[t]{0.48\textwidth}
        \centering
        \includegraphics[width=\linewidth]{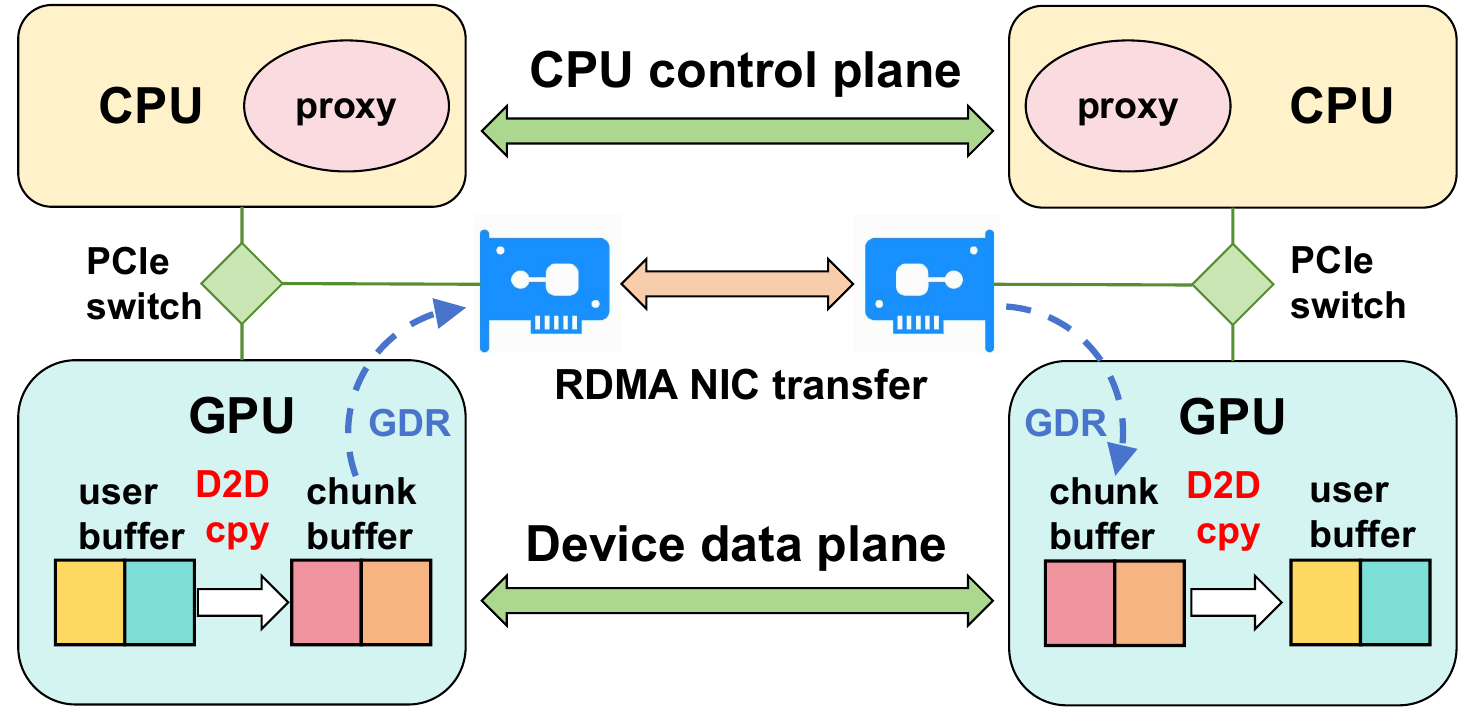}
        \caption{CPU-offloading P2P mechanism.}
        \label{fig:direct-path}
    \end{minipage}
    \hfill
    \begin{minipage}[t]{0.48\textwidth}
        \centering
        \includegraphics[width=\linewidth]{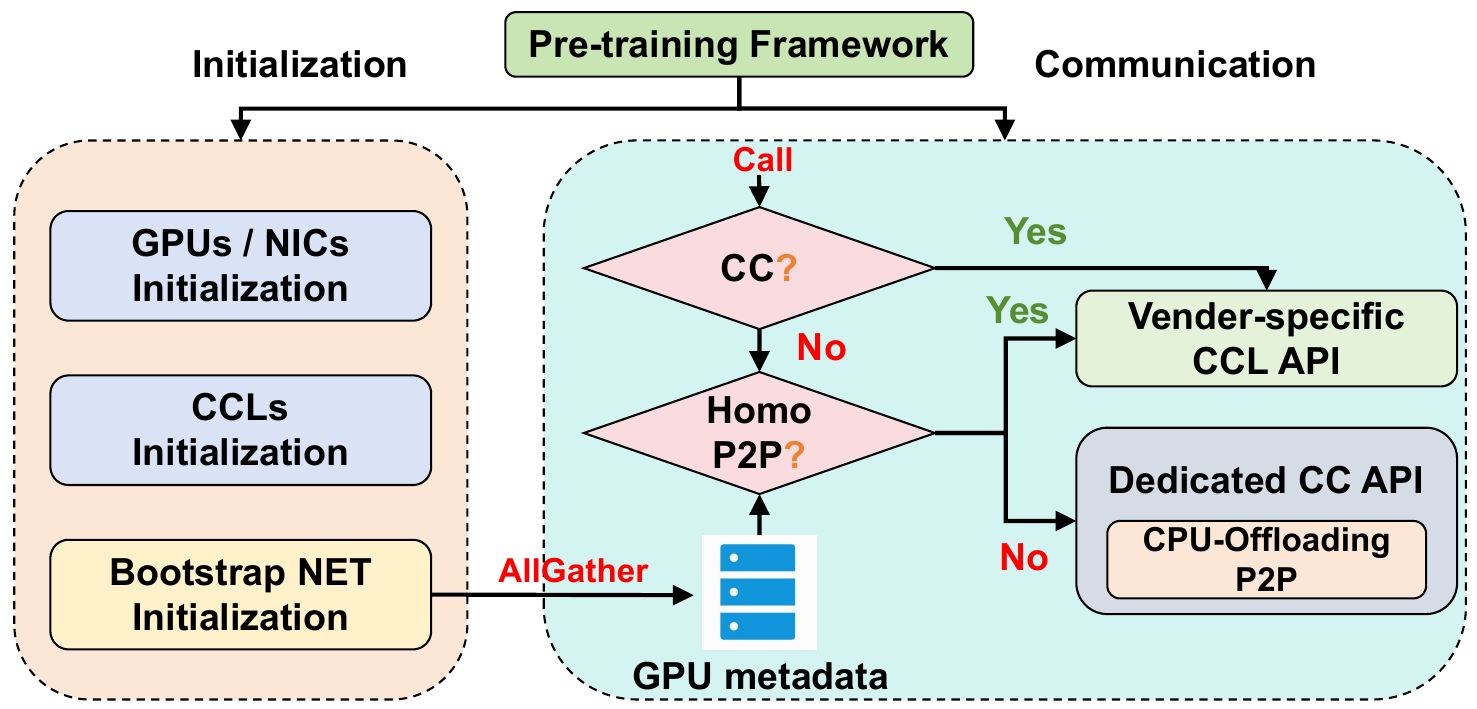}
        \caption{Heterogeneous mixed-training communication invocation workflow.}
        \label{fig:workflow-logic}
    \end{minipage}
\end{figure}

The architectural core of Device-Direct Communication is a CPU-offloading P2P mechanism (Figure \ref{fig:direct-path}): it decouples the control plane—including memory registration, connection management, and event handling—from the data plane. While the host manages control operations to maximize hardware compatibility, the data path remains entirely on-device, eliminating host-device memory copies overhead and mitigating PCIe bottlenecks. During a transfer, the source GPU first performs a device-to-device (D2D) memory copy to move data from the user buffer into chunk buffer, and then leverages GDR to transfer the data from the chunk buffer directly to NIC; once transmitted over the RDMA network, the receiver performs a symmetric operation to copy the data into GPU memory.

To abstract hardware heterogeneity, the approach adopts a multi-adaptor architecture consisting of Device, Net-Plugin, and CCL adaptors (refer to Table \ref{tab:adaptors} for functional details). For collective operations, each homogeneous subgroup (e.g., NVIDIA or AMD nodes) first performs intra-group aggregation using its vendor-specific CCL. Subsequently, the CPU-offloading P2P mechanism transfers these intermediate results across the heterogeneous cluster to achieve global collective communication synchronization. These capabilities were exposesed through a PyTorch backend plugin, allowing downstream applications (Megatron, DeepSpeed \cite{rasley2020deepspeed}) to invoke heterogeneous collectives seamlessly without altering high-level training logic.

As illustrated in Figure \ref{fig:workflow-logic}, this approach organizes the communication invocation workflow during in heterogeneous mixed pre-training into two primary stages:

\begin{enumerate}\item \textbf{Initialization Phase}\begin{itemize}\item \textbf{Resource Discovery}: Adaptors load vendor-specific interfaces for devices (e.g., GPUs and NICs) and CCLs, initialize hardware and CCL' s states, and establish the bootstrap network.\item \textbf{Topology Awareness}: Each rank executes an AllGather operation via the bootstrap network to synchronize and store the global cluster topology and GPU metadata.\end{itemize}\item \textbf{Communication Phase}
\begin{itemize}
    \item \textbf{Collective Communication (CC)}: Upon calling a collective operator, the system dispatches the task to the vendor-specific CCLs and orchestrates cross-cluster aggregation via the CPU-offloading P2P mechanism.
    \item \textbf{Point-to-Point (P2P) Communication}:
    \begin{itemize}
        \item \textbf{Homogeneous P2P}: Directly utilizes the vendor's native CCL P2P transport.
        \item \textbf{Heterogeneous P2P}: Facilitates direct data exchange between cross-vendor devices through dedicated CC interfaces, implemented via the CPU-offloading P2P mechanism.
    \end{itemize}
\end{itemize}
\end{enumerate}

By establishing these low-level communication links, the Device-Direct Communication significantly enhances the efficiency of heterogeneous LLM training at scale.

\begin{table}[t]
\centering
\caption{Classification and Functionality of Adaptors.}
\label{tab:adaptors}
\small
\renewcommand{\arraystretch}{1.5} % 增加行高，使垂直居中视觉效果更好
\begin{tabular}{@{} >{\centering\arraybackslash}m{2.5cm} >{\centering\arraybackslash}m{3cm} m{8.5cm} @{}} 
\toprule
\textbf{Adaptor} & \textbf{Functional Module} & \textbf{Core Implementation \& Technical Details} \\ 
\midrule
Device & Hardware Abstraction & Encapsulates vendor-specific GPU hardware interfaces (e.g., CUDA API for NVIDIA, ROCm API for AMD); provides unified primitives for memory allocation, stream synchronization, and event management. \\ 
\midrule
Net-Plugin & Network Adaptation & Abstracts hardware interfaces for various Network Interface Cards (NICs) via \textit{ibverbs}; handles RDMA memory region (MR) registration and queue pair management to facilitate high-speed data transfers. \\ 
\midrule
CCL & Collective Adaptation & Standardizes collective communication library interfaces across different vendors (e.g., NCCL, RCCL); supports optimized primitives such as \textit{AllReduce}, \textit{AllGather}, and \textit{ReduceScatter}. \\ 
\bottomrule
\end{tabular}
\end{table}

\section{Results and Analysis}
\label{result}
\subsection{Experimental Setup}

We evaluate the proposed heterogeneous mixed-training approach in terms of
\textbf{performance, stability, and correctness}, and compare its results across
\textbf{NVIDIA homogeneous}, \textbf{AMD homogeneous}, and
\textbf{AMD--NVIDIA heterogeneous} settings (In the following, unless otherwise noted, "approach" denotes Device-Direct Communication.). 

\subsubsection{Testbed}

All experiments are conducted using two physical nodes in total.

\begin{itemize}
  \item \textbf{NVIDIA node}: 8 $\times$ NVIDIA H200 GPUs \cite{h200}, interconnected via NVLink (900 GB/s bidirectional).
  \item \textbf{AMD node}: 8 $\times$ AMD Instinct MI325X GPUs \cite{mi325x}, interconnected via Infinity Fabric links \cite{infinity} (128 GB/s bidirectional).
  \item \textbf{Networking}: Each node is equipped with 8 $\times$ NVIDIA BlueField-3-E-Series-B3140H DPUs \cite{bluefield}, providing 100 GB/s bidirectional bandwidth per DPU.
\end{itemize}

The NVIDIA homogeneous and AMD homogeneous settings each consist
of a single node, where all GPUs are located within the same machine.
The heterogeneous setting is formed by connecting the NVIDIA and AMD
nodes, resulting in a two-node system.

\subsubsection{Workloads and Parallelization}

We use LLaMA-8B and Qwen2-7B as training workloads. The
parallel configuration is as follows:

\begin{itemize}
  \item Tensor Parallelism (TP) = 1
  \item Pipeline Parallelism (PP) = 2
  \item Data Parallelism (DP) = 4 for homogeneous settings
  \item Data Parallelism (DP) = 8 for the heterogeneous setting
\end{itemize}

To achieve optimal load balancing across pipeline stages and maximize per-GPU throughput, we tailor the layer partitioning strategies based on both model architectures and hardware configurations:
\begin{itemize}
\item \textbf{Homogeneous Setting}: For both NVIDIA and AMD clusters, LLaMA-8B is split evenly (16--16 layers), whereas Qwen2-7B follows a 15--13 layer distribution across the two pipeline stages.
\item \textbf{Heterogeneous Setting}: Layers are unevenly allocated to accommodate architectural differences:
\begin{itemize}
\item LLaMA-8B: 15 layers on AMD stage and 17 layers on NVIDIA stage.
\item Qwen2-7B: 12 layers on AMD stage and 16 layers on NVIDIA stage.
\end{itemize}
\end{itemize}

\subsection{Performance}
\label{sec:3-2}

\begin{figure}[t]
  \centering
  \includegraphics[width=1\linewidth]{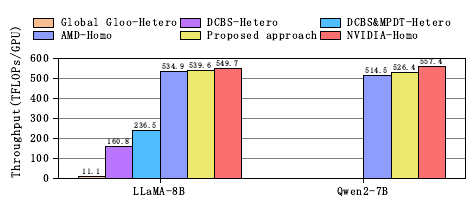}
  \caption{Average training throughput under different training configurations.}
  \label{fig:throughput}
\end{figure}

Figure \ref{fig:throughput} shows the average training throughput under different training configurations. For both LLaMA-8B and Qwen2-7B, the NVIDIA homogeneous configuration (NVIDIA-Homo) achieves the highest throughput, the AMD homogeneous configuration (AMD-Homo) exhibits the lower throughput, and the AMD–NVIDIA heterogeneous mixed-training configuration (proposed approach) falls between the two.

In the heterogeneous setting, the proposed approach achieves 98.2\% and 94.4\% of the throughput of the NVIDIA single-node baseline for LLaMA-8B and Qwen2-7B, respectively, while outperforming the AMD single-node configuration by 0.9\% and 2.3\%.

We further observe that, although DCBS\&MPDT optimizes the native Gloo backend in Megatron (Global Gloo-Hetero), its performance remains substantially inferior to that of the proposed Device-Direct Communication approach, primarily due to inefficient host–device memory copies. These results demonstrate that heterogeneous mixed training based on Device-Direct Communication can effectively aggregate computational resources across different GPU architectures and achieve performance close to that of a homogeneous NVIDIA system.

\subsection{Stability}
\begin{figure}[t]
  \centering
  \includegraphics[width=1\linewidth]{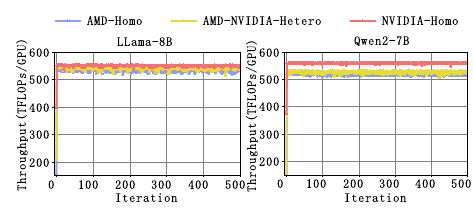}
  \caption{Throughput over 500 iterations of Llama-8B and Qwen2-7B.}
  \label{fig:throughput_curve}
\end{figure}

Figures \ref{fig:throughput_curve} show the throughput over 500 training iterations for LLaMA-8B and Qwen-7B, respectively, under the AMD homogeneous, NVIDIA homogeneous, and AMD–NVIDIA heterogeneous mixed-training configurations using Megatron. The throughput remains stable throughout the 500 iterations, exhibiting stability comparable to that of the homogeneous configurations, which demonstrates the robustness of the proposed heterogeneous mixed-training approach. In addition, the relative throughput ordering across different configurations is consistent with that observed in Fig.~\ref{fig:throughput}.

\begin{figure}[t]
  \centering
  \includegraphics[width=1\linewidth]{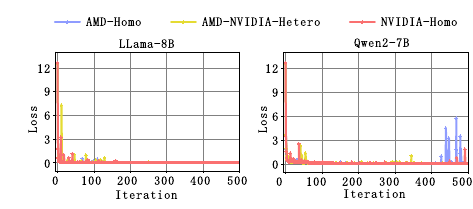}
  \caption{Loss over 500 iterations of Llama-8B and Qwen2-7B.}
  \label{fig:loss_curve}
\end{figure}

\subsection{Correctness}

Figure \ref{fig:loss_curve} shows the loss convergence curves of LLaMA-8B and Qwen-7B under different training configurations. The results indicate that the loss curves of the AMD–NVIDIA heterogeneous mixed-training configuration closely overlap with those of the AMD homogeneous and NVIDIA homogeneous baselines and exhibit consistent convergence behavior. This demonstrates that, under heterogeneous settings, the proposed Device-Direct Communication approach achieves numerical correctness and training dynamics equivalent to those observed in homogeneous settings.
\section{Discussion}
\label{discussion}

\subsection{Heterogeneity Limited to Pipeline Parallelism}

In this paper, heterogeneity is introduced only in the PP groups.
Although heterogeneous execution can in principle be supported in DP and TP groups, collective operations—including AllReduce and AllGather—are performed within these groups, and their performance is highly sensitive to the choice of collective algorithms and the characteristics of heterogeneous topologies. Given that the collective communication implementations of Device-Direct Communication approach are not tailored to any specific vendor hardware, extending heterogeneity to DP or TP groups may lead to unpredictable communication overhead. In contrast, PP involves only simple point-to-point communication and is therefore better suited for heterogeneous execution. For this reason, we restrict heterogeneity to PP.

\subsection{Effect of Model Partitioning}

The performance of heterogeneous mixed training strongly depends on an appropriate model partitioning strategy. With PP = 2, We adopt an uneven partitioning scheme, assigning fewer Transformer layers to the AMD GPUs due to their lower observed throughput, while allocating more Transformer layers to the NVIDIA GPUs. This design avoids pipeline imbalance and improves overall throughput. Under a proper partitioning strategy, the expected performance satisfies
\[
\text{AMD} < \text{AMD+NVIDIA} < \text{NVIDIA},
\]
which is consistent with our experimental results in section \ref{sec:3-2}. In contrast, improper partitioning can cause heterogeneous training to underperform even the slowest single-node configuration.

\subsection{Engineering Challenges}

Although the CCLs for AMD and NVIDIA GPUs (i.e., RCCL and NCCL) expose similar interfaces, enabling interoperable communication in heterogeneous mixed-training settings requires substantial engineering effort. Throughout development, training frequently encountered hang issues, which were resolved through extensive debugging and iterative engineering, ultimately achieving stable execution. This underscores that heterogeneous GPU training demands nontrivial system-level adaptation beyond nominal API-level compatibility.

\section{Conclusions}
\label{sec:conclusion}
This paper investigates heterogeneous mixed training in AMD–NVIDIA environments and presents a practical end-to-end training solution. We first implement a CPU-Forwarding Communication scheme as a compatibility-oriented baseline for heterogeneous execution. Building on this, we adopt a Device-Direct Communication approach that enables direct cross-vendor GPU communication and effectively eliminates the bottleneck of host-side memory copies. Experimental results on LLaMA-8B and Qwen2-7B demonstrate that the proposed Device-Direct Communication approach achieves up to 98\% of the throughput of an NVIDIA homogeneous system, while maintaining strong stability and correctness. This paper indicates that, with the proposed Device-Direct Communication approach, together with appropriate parallelization strategies and model partitioning, heterogeneous AMD–NVIDIA clusters can reliably support large-scale pre-training and efficiently utilize heterogeneous GPU resources.
%----------------------------------------------------------------------------------------
%	Bibliography
%----------------------------------------------------------------------------------------
\clearpage
\bibliography{bibliography/sample}{}

\end{document}